\documentclass[aps,showpacs,preprintnumbers,amsmath, amssymb]{revtex4}

\usepackage{float}
\usepackage{graphics,epsfig}
\usepackage{graphicx}
\usepackage{dcolumn}
\usepackage{bm}

\begin{document}
\baselineskip=0.8 cm

\title{{\bf No hair theorem for bound-state massless static scalar fields outside horizonless Neumann compact stars}}
\author{Yan Peng$^{1}$\footnote{yanpengphy@163.com}}
\affiliation{\\$^{1}$ School of Mathematical Sciences, Qufu Normal University, Qufu, Shandong 273165, China}

\vspace*{0.2cm}
\begin{abstract}
\baselineskip=0.6 cm
\begin{center}
{\bf Abstract}
\end{center}

We study no-hair theorem for horizonless objects, being subject to Neumann boundary conditions.
For massive scalar fields, a no hair theorem for Neumann compact stars was proved by us
in a previous paper, where the nonzero scalar field mass condition is essential in the proof.
In the present work, for massless scalar fields,
we prove a no hair theorem, which claims that bound-state massless static scalar fields
cannot exist outside asymptotically flat horizonless
neutral Neumann compact stars.

\end{abstract}

\maketitle
\newpage
\vspace*{0.2cm}

\section{Introduction}

Black holes are theoretical predictions of the general relativity gravity \cite{Einstein}.
And direct detections of gravitational waves provided mounting evidence that black holes indeed
exist in the universe \cite{BP}. Recently, the first ever image of a black hole has been captured by
a network of eight radio telescopes around the world \cite{picture}.
These discoveries open up hope to directly test various black hole theories from astronomical aspects.
One remarkable property of classical black holes is the famous no hair theorem \cite{Bekenstein}-\cite{JBN}.
If generically true, it would signify
that black hole solutions are very simple and uniquely
determined by three parameters: mass M, charge Q
and angular momentum J, see recent references
\cite{mr1}-\cite{JCE} and reviews \cite{Bekenstein-1,CAR}.

Interestingly, it was recently shown that such no hair theorem are not restricted
to black hole spacetimes. In the horizonless gravity,
Hod firstly proved no static scalar hair theorem
for asymptotically flat neutral reflecting compact stars \cite{Hod-6}.
When including a positive cosmological constant, it was found that
massive scalar, vector and tensor field hairs cannot exist
outside neutral horizonless reflecting compact stars \cite{Bhattacharjee}.
In the charged background, reflecting shells can exclude
scalar field hairs when the shell radius is large enough
\cite{Hod-8,Hod-9,Yan Peng-1}.
Large charged reflecting compact stars also cannot support the existence of scalar field hairs
\cite{Hod-10,Yan Peng-2,Yan Peng-3,Yan Peng-4,Rogatko2,Yan Peng-5}.
With nonminimal field-curvature couplings,
no hair theorem could still appear in the horizonless gravity \cite{nonm1,nonm2,nonm3}.
In addition, such no scalar hair theorem was further extended to horizonless compact
stars with Dirichlet surface boundary conditions \cite{Dirichlet}.

For massive scalar fields, no hair theorem was also proved for horizonless compact stars
with Neumann boundary conditions \cite{Yan Peng-6}.
The proof is based on the scalar field equation
$\psi\psi''+\frac{1}{2}(\frac{4}{r}+\nu'-\lambda')\psi\psi'-m^{2}e^{\lambda}\psi^2=0$,
where $\psi$ is the scalar field with nonzero mass m and $\nu$, $\lambda$
correspond to metric solutions.
At the extremum point of $\psi$,
there are relations $\psi\psi''\leqslant 0$, $\psi'=0$
and $m^{2}e^{\lambda}\psi^2<0$, which
are in contradiction with the scalar field equation.
This contradiction leads to the no hair theorem.
However, the nonzero scalar field mass assumption $m^2\neq 0$ is essential in the proof.
As is known, scalar field mass usually plays an important role in scalar condensations.
So it is meaningful to examine whether there is no hair theorem
for massless scalar fields in the background of Neumann compact stars.

In the following, we introduce the system of massless scalar fields
in the background of asymptotically flat neutral horizonless Neumann compact stars.
We prove a no hair theorem that bound-state massless scalar fields cannot exist outside
Neumann compact stars. We give conclusions in the last section.

\section{No massless scalar field hair outside Neumann compact stars}

We shall analyze the physical and mathematical properties of
an asymptotically flat gravity system, which is constructed by
a central compact object coupled to a static massless scalar field.
In Schwarzschild coordinates $(t,r,\theta,\phi)$, the line element of the
external spherically symmetric curved spacetime can be expressed in the form \cite{nonm1,nonm2}
\begin{eqnarray}\label{AdSBH}
ds^{2}&=&-ge^{-\chi}dt^{2}+\frac{dr^{2}}{g}+r^{2}(d\theta^2+sin^{2}\theta d\phi^{2}),
\end{eqnarray}
where $g$ and $\chi$ are functions only depending on the radial coordinate r.
In order to recover the flat spacetime at the infinity,
we impose asymptotical behaviors $g(r\rightarrow \infty)\backsim 1+O(r^{-1})$
and $\chi(r\rightarrow \infty)\backsim O(r^{-1})$.

The Lagrange density for a massless scalar field
in the asymptotically flat spacetime is
\begin{eqnarray}\label{lagrange-1}
\mathcal{L}=R-(\partial_{\alpha} \psi)^{2},
\end{eqnarray}
where R corresponds to the Ricci scalar curvature
and $\psi$ is the scalar field.

We consider the scalar field $\psi$ as a functions of r only. This leads
to the scalar field equation
\begin{eqnarray}\label{BHg}
\psi''+(\frac{2}{r}-\frac{\chi'}{2}+\frac{g'}{g})\psi'=0.
\end{eqnarray}

In the proof of the uniqueness of the Kerr solution
(Carter-Hawking-Robinson theorem),
metric solutions were assumed to be real analytic \cite{BCU,DRU,Hawking}.
In this work, we also assume that metric solutions are real analytic.
That is to say metric solutions can be locally expressed with power series.
We define the radial coordinate $r=r_{s}$ as the star surface radius.
For every $r_{\nu}\in[r_{s},R]$, there is a positive constant $\delta_{\nu}$
and in the range $(r_{\nu}-\delta_{\nu},r_{\nu}+\delta_{\nu})$,
metric solutions can be expanded with power series.
We pay attention to properties in the range $[r_{s},R]$.
We choose
\begin{eqnarray}\label{AdSBH}
(r_{1}-\delta_{1},r_{1}+\delta_{1}),(r_{2}-\delta_{2},r_{2}+\delta_{2}),\ldots,(r_{N}-\delta_{N},r_{N}+\delta_{N}),
\end{eqnarray}
which also satisfies
\begin{eqnarray}\label{AdSBH}
r_{1}=r_{s},~r_{N}=R~~and~~r_{i}<r_{i+1}-\delta_{i+1}<r_{i}+\delta_{i}<r_{i+1}~~for~~1\leqslant i\leqslant N-1.
\end{eqnarray}
So they combine to cover $[r_{s},R]$ as
\begin{eqnarray}\label{AdSBH}
[r_{s},R]\subset \bigcup_{i=1}^{i=N}(r_{i}-\delta_{i},r_{i}+\delta_{i}).
\end{eqnarray}

At the star surface, we impose the Neumann boundary condition
\begin{eqnarray}\label{AdSBH}
\psi'(r_{s})=0.
\end{eqnarray}

According to Cauchy-Kowalevski theorem,
in every range $(r_{i}-\delta_{i},r_{i}+\delta_{i})$,
solutions $\psi(r)$ of equation (3)
uniquely exist and can be expressed with power series \cite{BCU,DRU,Hawking,NAM}.
In the range $(r_{1}-\delta_{1},r_{1}+\delta_{1})$, the unique solution satisfying $\psi'(r_{s})=0$ is
\begin{eqnarray}\label{AdSBH}
\psi(r)=\psi(r_{s}).
\end{eqnarray}
In the range $(r_{2}-\delta_{2},r_{2}+\delta_{2})$, the scalar field can be
expressed as
\begin{eqnarray}\label{BHg}
\psi(r)=\sum_{n=0}^{\infty}a_{n}(r-r_{2})^{n}
\end{eqnarray}
with $a_{n}=\frac{\psi^{(n)}(r_{2})}{n!}$.
Since $(r_{2}-\delta_{2},r_{1}+\delta_{1})\subset(r_{1}-\delta_{1},r_{1}+\delta_{1})$,
we conclude that $\psi(r)$ is a constant $\psi(r_{s})$ in $(r_{2}-\delta_{2},r_{1}+\delta_{1})$ according to (8).
Relation (5) yields $(r_{2}-\delta_{2},r_{1}+\delta_{1})\subset (r_{2}-\delta_{2},r_{2}+\delta_{2})$.
So $\psi(r)$ can be expressed as (9) in $(r_{2}-\delta_{2},r_{1}+\delta_{1})$.
Also considering that (9) is a constant in $(r_{2}-\delta_{2},r_{1}+\delta_{1})$,
we deduce relations
\begin{eqnarray}\label{BHg}
a_{0}=\psi(r_{s}),
\end{eqnarray}
\begin{eqnarray}\label{BHg}
a_{n}=0~~~for~~all~~~n\geqslant 1.
\end{eqnarray}
Then in the larger range $(r_{2}-\delta_{2},r_{2}+\delta_{2})$, there is
\begin{eqnarray}\label{BHg}
\psi(r)=a_{0}=\psi(r_{s}).
\end{eqnarray}

Following this analysis, in the range $(r_{3}-\delta_{3},r_{3}+\delta_{3})$, the scalar field can be
expressed as
\begin{eqnarray}\label{BHg}
\psi(r)=\sum_{n=0}^{\infty}b_{n}(r-r_{3})^{n}
\end{eqnarray}
with $b_{n}=\frac{\psi^{(n)}(r_{3})}{n!}$.
Since $(r_{3}-\delta_{3},r_{2}+\delta_{2})\subset(r_{2}-\delta_{2},r_{2}+\delta_{2})$,
we deduce that $\psi(r)$ is a constant $\psi(r_{s})$ in $(r_{3}-\delta_{3},r_{2}+\delta_{2})$ according to (12).
As (13) is a constant $\psi(r_{s})$ in $(r_{3}-\delta_{3},r_{2}+\delta_{2})$, there is
\begin{eqnarray}\label{BHg}
b_{0}=\psi(r_{s}),
\end{eqnarray}
\begin{eqnarray}\label{BHg}
b_{n}=0~~~for~~all~~~n\geqslant 1.
\end{eqnarray}
Then in the range $(r_{3}-\delta_{3},r_{3}+\delta_{3})$, there is
\begin{eqnarray}\label{BHg}
\psi(r)=b_{0}=\psi(r_{s}).
\end{eqnarray}

Following this method, we can further obtain $\psi(r)=\psi(r_{s})$
in other ranges $(r_{4}-\delta_{4},r_{4}+\delta_{4})$,
$(r_{5}-\delta_{5},r_{5}+\delta_{5})$,\ldots,$(r_{N}-\delta_{N},r_{N}+\delta_{N})$.
So in the range $[r_{s},R]\subset \bigcup_{i=1}^{i=N}(r_{i}-\delta_{i},r_{i}+\delta_{i})$, the scalar field solution is
\begin{eqnarray}\label{BHg}
\psi(r)=\psi(r_{s}).
\end{eqnarray}
In particular, (17) yields the relation
\begin{eqnarray}\label{BHg}
\psi(R)=\psi(r_{s}).
\end{eqnarray}

In the far region, the scalar field equation can be
approximated by the differential equation
\begin{eqnarray}\label{BHg}
\psi''+\frac{2}{r}\psi'=0,
\end{eqnarray}
whose general solution is
\begin{eqnarray}\label{AdSBH}
\psi\sim A+\frac{B}{r}
\end{eqnarray}
with A and B as integral constants.
In this work, we study the bound-state scalar configurations \cite{bound}.
It means that the scalar field must asymptotically approaches
zero at the infinity. So we fix $A=0$. Then
the scalar field satisfies the asymptotical behavior
\begin{eqnarray}\label{AdSBH}
\psi\sim \frac{B}{r}.
\end{eqnarray}
We can choose a large enough R satisfying
\begin{eqnarray}\label{BHg}
\psi(R)<\psi(r_{s}).
\end{eqnarray}
Relation (22) is in contradiction with the relation (18),
which means that the equation (3) cannot possess no nontrivial solutions.
So we conclude that bound-state massless scalar field hairs cannot exist outside
neutral horizonless Neumann compact stars in the
asymptotically flat spherical background.

\section{Conclusions}

We studied no hair theorem for bound-state static massless scalar fields
outside neutral horizonless Neumann compact stars.
We chose the asymptotically flat spherically
symmetric spacetime. We found that the solution should satisfy (18),
which is in contradiction with the relation (22).
This contradiction leads to the fact that no nontrivial
scalar field solution of equation (3) cannot exist.
We concluded that neutral horizonless Neumann compact stars
cannot support the bound-state massless scalar field hair
in the asymptotically flat spherically symmetric background.

\begin{acknowledgments}

We would like to thank the anonymous referee for the constructive suggestions to improve the manuscript.
This work was supported by the Shandong Provincial Natural Science Foundation of China under Grant
No. ZR2018QA008. This work was also supported by a grant from Qufu Normal University of China under
Grant No. xkjjc201906.

\end{acknowledgments}

\end{document}